\documentclass{article}

\usepackage{microtype}
\usepackage{graphicx}
\usepackage{subfigure}
\usepackage{booktabs} 
\usepackage{multirow}

\usepackage{hyperref}

\usepackage[accepted]{icml2023}

\usepackage{amsmath}
\usepackage{amssymb}
\usepackage{mathtools}
\usepackage{amsthm}

\usepackage[capitalize,noabbrev]{cleveref}

\theoremstyle{plain}

\theoremstyle{definition}

\theoremstyle{remark}

\usepackage[textsize=tiny]{todonotes}

\icmltitlerunning{Re-Dock: Towards Flexible and Realistic Molecular Docking with Diffusion Bridge}

\begin{document}

\twocolumn[
\icmltitle{Re-Dock: Towards Flexible and Realistic Molecular\\ Docking with Diffusion Bridge}



\icmlsetsymbol{equal}{*}

\begin{icmlauthorlist}
\icmlauthor{Yufei Huang}{equal,zju,westlake}
\icmlauthor{Odin Zhang}{equal,zju,uw}
\icmlauthor{Lirong Wu}{zju,westlake}
\icmlauthor{Cheng Tan}{zju,westlake}
\icmlauthor{Haitao Lin}{zju,westlake}

\icmlauthor{Zhangyang Gao}{zju,westlake}
\icmlauthor{Siyuan Li}{zju,westlake}
\icmlauthor{Stan. Z. Li}{westlake}
\end{icmlauthorlist}

\icmlaffiliation{zju}{Zhejiang University, Hangzhou}
\icmlaffiliation{westlake}{AI Lab, Research Center for Industries of the Future, Westlake University}
\icmlaffiliation{uw}{University of Washington}

\icmlcorrespondingauthor{Stan Z. Li}{cairi@westlake.edu.cn}

\icmlkeywords{Machine Learning, ICML}

\vskip 0.3in
]



\printAffiliationsAndNotice{\icmlEqualContribution} 

\begin{abstract}
Accurate prediction of protein-ligand binding structures, a task known as molecular docking is crucial for drug design but remains challenging. While deep learning has shown promise, existing methods often depend on holo-protein structures (\textit{docked}, and not accessible in \textit{realistic} tasks) or neglect pocket sidechain conformations, leading to limited practical utility and unrealistic conformation predictions. To fill these gaps, we introduce an under-explored task, named \textit{flexible docking} to predict poses of ligand and pocket sidechains simultaneously and introduce \texttt{Re-Dock}, a novel diffusion bridge generative model extended to geometric manifolds. Specifically, we propose energy-to-geometry mapping inspired by the Newton-Euler equation to co-model the binding energy and conformations for reflecting the energy-constrained docking generative process. Comprehensive experiments on designed benchmark datasets including apo-dock and cross-dock demonstrate our model's superior effectiveness and efficiency over current methods.
\end{abstract}
\vspace{-2em}
\section{Introduction}
\label{sec1}
Proteins can have their biological functions~\cite{huang2023dataefficient, huang2023protein, wu2022survey} altered by the binding of small molecule ligands, such as drugs~\cite{lin2022diffbp, lin2023functional,hu2022protein}. Molecular docking which reveals this interaction, is critical in drug design and involves predicting the conformation of a protein-ligand complex. A key challenge in molecular docking lies in the \textit{induced-fit} mechanism~\cite{sherman2006novel}, where the protein's binding sites (pockets) are \textit{flexible}, altering their poses in response to ligand binding. Notably, the pocket's sidechain atoms exhibit the most significant flexibility~\cite{clark2019inherent}.
\begin{figure}[!tbp]
    \begin{center}
    \hspace{-1em}
        \subfigure[Flexible docking and \textit{priori} leakage]{\includegraphics[width=0.7\linewidth]{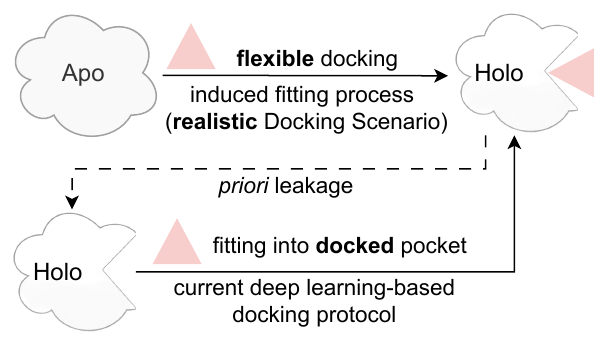}\label{fig:1a}}
        \subfigure[Steric clashes]{\includegraphics[width=0.3\linewidth]{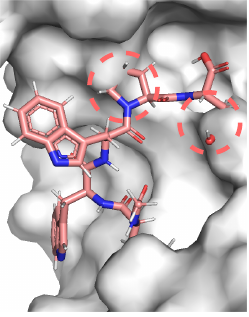}\label{fig:1b}}
    \end{center}
    \vspace{-1.7em}
    \caption{The illustration of our motivation. (a) The dotted line represents current protocols take docked pockets as input, which are not accessible in realistic scenarios and provide hints or leakage for ligand poses prediction. (b) The result of DiffDock (pdb\_id: 6nsv); the steric clashes are highlighted with red circles, where the ligand overlaps with the protein surface (i.e., sidechains).}
    \label{fig:1}
    \vspace{-1.8em}
\end{figure}

Though important, accurately predicting the bound structures is highly challenging. Traditional docking methods~\cite{autodock2021,alhossary2015qvina}, using empirical scoring functions and optimization algorithms, struggle with the vast search space and complex calculations, often resulting in inaccurate and slow predictions. Recent deep learning approaches~\cite{zhou2023unimol,liao2019deepdock} focus on predicting ligand binding poses using bound protein structures (holo-structures), which include ground truth sidechain conformations (i.e. \textit{priori} leakage). These approaches, by introducing \textbf{\textit{priori} leakage} (as illustrated in Fig. \ref{fig:1a}), oversimplify the problem and fail to mimic realistic docking scenarios. Other deep learning methods~\cite{pei2023fabind, corso2023diffdock} ignore sidechains for simplicity and implicit flexibility. \textbf{This neglect often results in unrealistic poses} where ligands overlap with side chains, i.e. steric clashes (as shown in Fig.\ref{fig:1b}). More accurate and realistic binding pose predictions require explicit sidechain modeling. Additionally, protein-molecule interaction is the key to model the binding process. Current approaches learn this interaction implicitly with neural networks while \textbf{fail to incorporate explicit modeling of interaction} in 3D coordinates for direct and accurate guidance.

To address these limitations, we introduce an under-explored task, named \textit{flexible docking}~\cite{sahu2024review} to predict poses of ligand and pocket sidechains simultaneously and introduce \texttt{Re-Dock}, a flexible and \underline{Re}alistic generative \underline{Dock}ing framework with explicit modeling of pocket sidechain flexibility and integrated interaction prior to steer the generation process. \texttt{ReDock} mimics the induced-fit process~\cite{sherman2006novel} for \textit{realistic docking scenarios} and generates \textit{physically realistic conformations} by extending diffusion bridges~\cite{priorbridge} to non-Euclidean manifolds of \textit{implicit} geometries: rotations, translations, and torsion with \textit{explicit} interaction prior in Euclidean space. 

In detail, we construct neural diffusion models to imitate the bridge processes for generating flexible and realistic poses of both ligands and pockets. Diffusion bridges are stochastic processes that guarantee to yield given observations at the fixed terminal time~\cite{liu2022let}. Notably, we model sidechain distributions autoregressively for better generation quality concerning their sequential nature~\cite{zhang2023diffpack}. Unlike previous diffusion bridge processes defined over Euclidean space with molecular coordinates~\cite{priorbridge}, we explore bridge processes in geometric space, challenging for its implicit nature of modeling data points. For constructing interaction-informed prior bridges over geometry, we enable \textit{Energy-to-Geometry mapping} using the Newton-Euler equations inspired by rigid body mechanics.

We benchmark \textit{flexible docking} in the pocket-aware setting and provide generalized results of \texttt{Re-Dock} with pocket predictions. In many drug discovery pipelines, pockets are identified early~\cite{zhang2023learning} and their sidechains contribute to the majority of flexibility~\cite{clark2019inherent} during docking. Thus \texttt{Re-Dock} focuses on these flexible pocket sidechains. We design a new benchmark that reflects realistic scenarios, including apo crystal docking and cross-dock using the PDBBind~\cite{liu2017PDBBind} and our curated datasets. \texttt{Re-Dock} shows comparative results in overall benchmark tasks, proving its effectiveness in predicting flexible docking structures. This demonstrates its potential for real-world applications. Our key contributions are:
\begin{itemize}
\vspace{-1em}
\item We introduce the under-explored task of \textit{flexible docking} and design a rigorous benchmark with new datasets.
\vspace{-1.8em}
\item We propose \texttt{Re-Dock}, a novel diffusion bridge model extended to non-Euclidean manifolds with \textit{energy-to-geometry mapping} inspired by mechanics. It enables interaction-aware, `induced', generative docking processes with co-modeling of binding energy and poses.
\vspace{-1.8em}
\item Superior benchmark test results including cross-dock suggest our potential for real-world applications. 
\end{itemize}

\section{Related Works}
\noindent \textbf{Molecular Docking.} This field predicts how proteins and ligands bind together. Traditional approaches, like AutoDock Vina~\cite{autodock2021, alhossary2015qvina}, SMINA~\cite{koes2013smina}, and GLIDE~\cite{halgren2004glide}, utilize energy-based functions. Recently, deep learning, especially Graph Neural Network~\cite{NEURIPS2022_KRD, Wu2023Homo, zheng2023lightweight} has brought two innovations: Regression-based methods such as Equibind~\cite{stärk2022equibind}, Tankbind~\cite{Lu2022TankBind}, and Fabind~\cite{pei2023fabind} for predicting the ligand's docking pose directly and generative docking introduced by DiffDock~\cite{corso2023diffdock,plainer2023diffdock_pocket} approaching docking as the generation of ligand geometries, like rotations. Most methods assume known holo-protein structures, which is often impractical. Traditional approaches and some structure prediction methods~\cite{qiao2023state, krishna2023generalized} can account for protein flexibility but require extensive computations. Our work lies in generative docking with pocket flexibility and explicit modeling of interaction, thus can be applied to more realistic tasks, including cross-dock. More related works can be found in the Appendix. A.

\noindent \textbf{Diffusion Bridge Process.} Diffusion-based generative models~\cite{song2020score, ho2020denoising} have become popular in AI generation by adding noise to data and then using a reverse process to generate outputs. Efforts to improve these models have led to diffusion bridges. Schrodinger bridges~\cite{BortoliDSB2021,shi2023diffusion} have been proposed for learning entropy-regularized optimal transports of generation paths and guarantee desirable outputs, but these models involve iterative proportional fittings and are computationally costly. Research works such as \cite{peluchetti2021non} and prior bridge~\cite{priorbridge} directly learn diffusion trajectories with specific ending data points and inject problem-dependent prior into these paths, avoiding the time-reversal technique of \cite{song2020score}. However, applying diffusion bridges to geometric domains (e.g., rotations and torsion) requires designing appropriate bridge processes, which is an area that remains unexplored.

\begin{figure*}[!tbp]
    \begin{center}
        \includegraphics[width=1.0\linewidth]{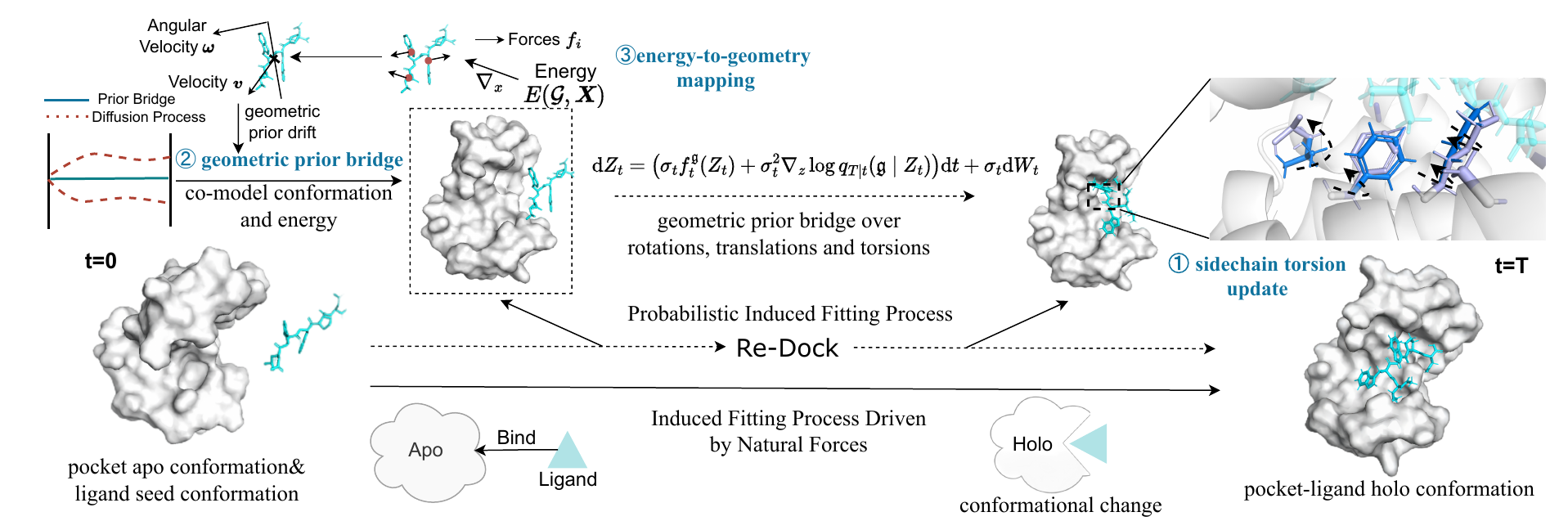}
    \end{center}
    \vspace{-1em}
    \caption{The illustration of \texttt{Re-Dock} Framework. We aim to simulate the induced fitting process with geometric prior bridges. Our key designs are threefold: \raisebox{.5pt}{\textcircled{\raisebox{-.9pt} {1}}} The pocket sidechains displace the most flexibility for inducing interactions. Thus, we generate the sidechain conformations (the blue and purple sticks are the conformations of two steps before and after, respectively; we omit other sidechains for simplicity) via torsion angle updates while docking. \raisebox{.5pt}{\textcircled{\raisebox{-.9pt} {2}}} We explore a novel generative model, the geometric prior bridge for reflecting the energy-constrained fitting process. Compared with the diffusion processes (the red curves), the prior bridge process (the blue line) is augmented with problem-dependent prior and thus more fast and accurate to generate.
    \raisebox{.5pt}{\textcircled{\raisebox{-.9pt} {3}}} For explicit modeling of interaction and constructing prior bridges over geometries, we propose an energy-to-geometry mapping module inspired by Newton-Euler equations.}
    \label{fig:2}
    \vspace{-0.8em}
\end{figure*}

\section{Backgrounds}
\label{sec3}
\subsection{Diffusion Bridge for Non-geometric domains}
\label{sec3.1}
In this section, we provide an overview of the general diffusion bridge to introduce the necessary notations and concepts based on ~\cite{priorbridge,liu2022let}. 

\noindent \textbf{Definition.} Diffusion bridges are diffusion processes that are conditioned to initialize and terminate at two given states. We can learn a generative model given a dataset $\{x^{(k)}\}^n_{k=1}$ drawn from an unknown distribution $\Pi^{*}$ on $\mathbb{R}^d$ via building and imitating such a diffusion bridge with prior and data distribution on both sides. A diffusion bridge generative model on the time interval [0,1] is:
\vspace{-0.4em}
\begin{equation}
\begin{small}
\mathbb{P}^\theta{:}\quad\mathrm{d}Z_t=s_t^\theta(Z_t)\mathrm{d}t+\sigma_t(Z_t)\mathrm{d}W_t, \forall t\in[0,1], Z_0\sim\mu_0.
\label{equ:0}
\end{small}
\end{equation}
where $W_t$ is a standard Wiener process; $\sigma_t: \mathbb{R}^d \rightarrow \mathbb{R}^{d \times d}$ is a positive definition covariance coefficient; $s_t^\theta: \mathbb{R}^d \rightarrow \mathbb{R}^d$ is parameterized as a neural network with parameter $\theta$, and $\mu_0$ is the prior distribution. Here, $\mathbb{P}^\theta$ represents the distribution of the diffusion trajectory $Z=\{Z_t:t\in[0,1]\}$ and $\mathbb{P}_t^\theta$ is the marginal distribution of $Z_t$ at time $t$. We aim at learning a generative model with parameter $\theta$ such that the distribution $\mathbb{P}^\theta_1$ of the terminal state $Z_1$ equals the data distribution $\Pi^{*}$.

\noindent \textbf{Building and Learning Diffusion Bridges.} There are infinitive diffusion processes $\mathbb{P}^\theta$ that reach the same terminal distribution but differ in distributions of trajectories $Z$. Diffusion Bridge has the appealing potential to inject problem-dependent prior information into the trajectories and learning processes to obtain a generative model $\mathbb{P}^\theta$ fast and accurately. To achieve this, \cite{priorbridge} elicit an \textit{imputation} process $\mathbb{Q}^x$ for each $x \in \mathbb{R}^d$, such that a draw $Z \in \mathbb{Q}^x$ yields trajectories that 1) are pinned at $Z_1=x$, and 2) reflect important physical prior information on the current problem with more physically regularized processes.

Formally, if $\mathbb{Q}^x(Z_1=x)=1$, then $\mathbb{Q}^x$ or simply an $x$-bridge is a bridge process ending at data point $x$. We can construct physically informed diffusion bridges based on $\mathbb{Q}^x$ if we first sample a data point $x\in\Pi^*$ and then draw a bridge $Z\in\mathbb{Q}^x$ pinned at $x$. Thus the distribution of trajectories $Z$ is a mixture of $\mathbb{Q}^x$: ${\mathbb{Q}^{\Pi^*}}:=\int\mathbb{Q}^x(\cdot)\Pi^*(\mathrm{d}x)$. 

We can learn the diffusion bridge model $\mathbb{P}^\theta$ by imitating the trajectories drawn from $\mathbb{Q}^{\Pi^*}$ since the crucial property of $\mathbb{Q}^{\Pi^*}$ is its end distribution equals the data distribution, i.e., $\mathbb{Q}^{\Pi^*}_1=\Pi^*$. This can be formulated by maximum likelihood or equivalently minimizing the KL divergence: $\min_\theta\left\{\mathcal{L}(\theta):=KL(\mathbb{Q}^{\Pi^*}\mid\mid\mathbb{P}^\theta)\right\}$. Furthermore, assume that the bridge $\mathbb{Q}^x$ is a diffusion model of form:
\vspace{-0.3em}
\begin{equation}
\begin{small}
\mathbb{Q}^x{:}\quad\mathrm{d}Z_t=b_t(Z_t\mid x)\mathrm{d}t+\sigma_t(Z_t)\mathrm{d}W_t,Z_0\sim\mu_0.
\label{equ:1}
\end{small}
\vspace{-0.3em}
\end{equation}
where $b_t(Z_t|x)$ is an $x$-dependent drift term that needs to be carefully designed to meet the bridge condition and incorporate prior information simultaneously. We adopt a practical and simple family of bridges proposed by \cite{priorbridge} in this paper by introducing modifications to Brownian Bridges~\cite{liu2022let}: 
\vspace{-0.3em}
\begin{equation}
\begin{small}
\hspace{-0.5em}
\mathbb{Q}^x_{\mathrm{bb},f}:\mathrm{d}Z_t=\left(\sigma_tf_t(Z_t)+\sigma_t^2\frac{x-Z_t}{\beta_1-\beta_t}\right)\mathrm{d}t+\sigma_t\mathrm{d}W_t.
\label{equ:2}
\end{small}
\vspace{-0.3em}
\end{equation}
where $\beta_t=\int_0^t\sigma_s^2\mathrm{d}s$, and $f_t(Z_t)$ is an extra drift term which reflects physical prior (e.g., the physical force) and $Z_0\in\mathcal{N}(x,\beta_1)$ or $\mathcal{N}(0,\beta_1)$ when $\beta_1$ is large enough~\cite{liu2022let}.  
After choosing appropriate $\mathbb{Q}^x_{bb,f}$, using Girsanov theorem, the loss function can be reformed into a form of denoising score matching loss of~\cite{heng2022simulating}:
\vspace{-0.5em}
\begin{equation}
\begin{aligned}
\mathcal{L}(\theta) = \mathbb{E}_{Z\sim\mathbb{Q}^{\Pi^*}}\Bigl[\frac{1}{2}\int_0^1 \Bigl\| \sigma(Z_t)^{-1}&(s_t^\theta(Z_t) \Bigr. \Bigr.\\
\Bigl. \Bigl. - b_t(Z_t\mid Z_1&))\Bigr\|_2^2 \, \mathrm{d}t\Bigr] + \mathrm{const}. 
\label{equ:3}
\end{aligned}
\end{equation}
where the const term contains the log-likelihood for the initial distribution $\mu_0$, which is a const in our problem. 
\vspace{-1em}
\subsection{Problem Statement}
\label{sec3.2}
\vspace{-0.5em}
Given unbounded protein and ligand structures as inputs, \textit{flexible docking} aims to predict the binding pose (i.e. atom coordinates) of the pocket sidechain and ligand, imitating the process of induced-fit. For simplicity and efficiency, we model the \textit{changes} the ligand and sidechains undergo during binding following \cite{corso2023diffdock}, instead of directly modeling coordinates. Thus, we focus on an $(m+6)$-dimensional manifold, where $m$ represents the number of rotatable bonds, and six extra dimensions account for roto-translations of ligands relative to the fixed protein backbone. 

Our approach differs from previous methods as we construct diffusion bridges $\mathbb{P}^\theta$ on geometric manifolds (e.g. rotation $SO(3)$ and torsion $SO(2)^m$) and integrate interaction priors into our \textit{generative bridge training}. Next, we detail how to learn the probabilistic induced-fit process $\mathbb{P}^\theta$.



\vspace{-1em}
\section{Method}
In this section, we formally describe the flexible and \underline{Re}alistic protein-ligand \underline{Dock}ing (Re-Dock) framework. 
We aim to model protein-ligand interactions more accurately by considering the flexibility of pocket sidechains and incorporating a novel geometric diffusion bridge model to inject interaction prior. 
Our method draws inspiration from recent advances in Diffusion Bridge~\cite{liu2022let,priorbridge}, which integrates problem-dependent prior into generative paths. However, constructing suitable bridge processes on geometry with interaction prior is challenging. We address this by extending the diffusion bridge to geometric sub-manifolds and incorporating interaction prior with the Newton-Euler Equation. The overall framework is introduced in Section.\ref{sec4.1}, with further elaboration on details of our method and its learning process with energy-conformation co-modeling in Section.\ref{sec4.2} and Section.\ref{sec4.3}, respectively. A high-level schematic is provided in Fig. \ref{fig:2}.
\vspace{-1em}
\subsection{Diffusion Prior Bridge on Geometries}
\label{sec4.1}
The geometric sub-manifold $\mathcal{M}_C$ is a product space $\mathbb{G}$ of the 3D translation group $\mathbb{T}(3)$, the 3D rotation group $SO(3)$ of rigid rotations of the ligand, and the 2D rotation groups $SO(2)^m$ of changes in m torsion angles $\theta$ of the pocket-ligand complex. As a starting point, We first build a simple Brownian Bridge on $\mathbb{G}$ via Doob's h-transforms~\cite{liu2022let}. It states the Brownian bridge $\mathbb{Q}^x_{bb}$ can be shown to be the law of:
\vspace{-0.5em}
\begin{equation}
\begin{small}
\hspace{-0.5em}
\mathbb{Q}^x_{bb}:\mathrm{d}Z_t=\sigma_t^2\nabla_z\log q_{T|t}(x\mid Z_t)\mathrm{d}t+\sigma_t\mathrm{d}W_t.
\label{equ:4}
\end{small}
\end{equation}
where $q_{T|t}(x\mid Z_t)$ is the density function of the transition probability $\mathbb{Q}_{T|t}(\cdot|z)=\mathcal{N}(z,\beta_{T}-\beta_{t})$,where $\beta_{t}=\int_{0}^{t}\sigma^{2}\mathrm{d}s$.
Since $\mathbb{G}$ is a product manifold, the forward diffusion proceeds independently in each manifold, and the tangent space is a direct sum: $T_{g}\mathbb{Q}=T_{\mathbf{r}}\mathbb{T}_{3}\oplus T_{R}SO(3)\oplus T_{\boldsymbol{\theta}}SO(2)^{m}\cong\mathbb{R}^{3}\oplus\mathbb{R}^{3}\oplus\mathbb{R}^{m}$ where g = ($\mathbf{r}, R, \boldsymbol{\theta}$). Thus, the construction of the bridge on $\mathbb{G}$ is equivalent to building Brownian Bridges on each geometry independently.
In all three groups, we can define the Brownian Bridge as Equation~\ref{equ:4} where $\sigma = \sigma^{\mathbf{r}}, \sigma^R, \sigma^\theta$ for $\mathbb{T}(3), SO(3), SO(2)^m$ respectively, $\mathcal{N}(\cdot,\cdot)$ is the corresponding \textit{normal distribution} on geometry and $\mathrm{W}$ is the corresponding Brownian motion. As mentioned in Section.\ref{sec3.1}, we can add an extra drift term $f_t(z_t)$ that reflects the interaction prior and design a prior bridge $\mathbb{Q}_{bb,f}^{\mathfrak{g}}$ where $\mathfrak{g}=\mathbf{r},R,\boldsymbol{\theta}$ as:
\vspace{-0.5em}
\begin{equation}
\begin{small}
\mathrm{d}Z_t=\left(\sigma_tf_t^\mathfrak{g}(Z_t) + \sigma_t^2\nabla_z\log q_{T|t}(\mathfrak{g}\mid Z_t)\right)\mathrm{d}t+\sigma_t\mathrm{d}W_t.
\label{equ:5}
\end{small}
\end{equation}
Since $\mathbb{T}(3)\cong\mathbb{R}^3$, it is trivial to build the bridge on translation in the same form of Equation~\ref{equ:2} with variance $\sigma_t^r$. The $SO(2)^m$ group is diffeomorphic to the torus $\mathbb{T}^m$, on which the diffusion kernel is a \textit{wrapped normal distribution}~\cite{jing2022torsional} with variance schedule $\sigma_t^\theta$. The transition probability $q_{T|t}(\theta\mid \theta_t)$ is:
\vspace{-1em}
\begin{equation}
\begin{small}
\hspace{-1em}
q_{T|t}(\boldsymbol{\theta}|\boldsymbol{\theta}_t)\propto\sum_{\boldsymbol{d}\in\mathbb{Z}^m}\exp\left(-\frac{\|\boldsymbol{\theta}-\boldsymbol{\theta}^{\prime}+2\pi\boldsymbol{d}\|^2}{\beta_T-\beta_t}\right)
\label{equ:6}
\end{small}
\end{equation}
This can be sampled directly, and the score can be precomputed as a truncated infinite series. We consider the isotropic Gaussian distribution on SO(3)~\cite{leach2022denoising}. 
The transition kernel on $SO(3)$ is $\mathcal{IG}_{SO(3)}(R, \beta_T-\beta_t)$ distribution, which can be sampled in the axis-angle parameterization by sampling a unit vector $\hat{\omega}\in\mathfrak{so}(3)$ uniformly and random angle $\omega \in [0, \pi]$ according to:
\vspace{-1em}
\begin{equation}
\hspace{-2em}
\begin{small}
\begin{aligned}
p(\omega)&=\frac{1-\cos\omega}\pi f(\omega), \sigma=\beta_T-\beta_t,\\
\quad f(\omega)&=\sum_{l=0}^\infty(2l+1)\exp(-l(l+1)\sigma^2/2)\frac{\sin((l+1/2)\sigma^2}{\sin(\omega/2)}
\label{equ:7}
\end{aligned}
\end{small}
\vspace{-0.5em}
\end{equation}
The score computation and sampling can be accomplished efficiently by precomputing the truncated infinite
series and interpolating the CDF of $p(\omega)$, respectively.

\begin{figure}[!tbp]
    \begin{center}
    \hspace{-1em}
        \subfigure[Sidechain Angles]{\includegraphics[width=0.43\linewidth]{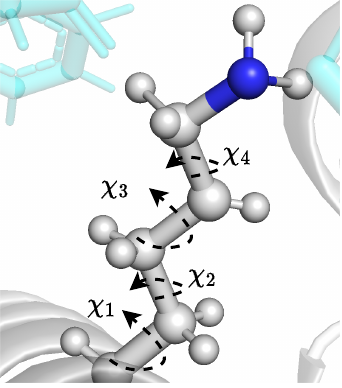}\label{fig:3a}}
        \subfigure[Effect of rotating $\theta^{sc}_1$]{\includegraphics[width=0.55\linewidth]{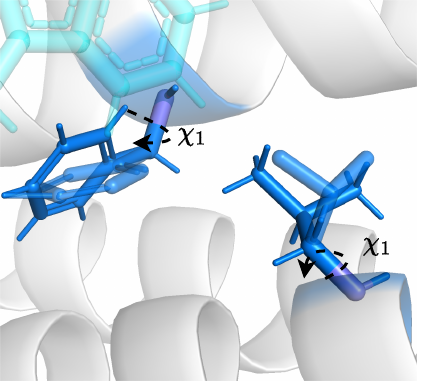}\label{fig:3b}}
    \end{center}
    \vspace{-1.5em}
    \caption{The illustration of sidechain updates. (a) Up to four sidechain $\theta^{sc}$ angles (formally, $\chi$ angles) have a sequential order. (b) Rotating $\theta^{sc}_1$ will affect the coordinates of atoms in $\theta^{sc}_2$, $\theta^{sc}_3$ and $\theta^{sc}_4$. It's similar for rotating $\theta^{sc}_2$ and $\theta^{sc}_3$. The atom groups of the later angles will accumulate noise from the former angles which complicates the latter's denoising process. }
    \label{fig:3}
    \vspace{-1.5em}
\end{figure}

\noindent \textbf{Autoregressive update on sidechains.}
Adding noises to (i.e. rotating) one angle $\boldsymbol{\theta}^{sc}_i$ will result in changes in and beyond the corresponding atom groups (for example, rotating $\theta^{sc}_1$ as in Fig. \ref{fig:3b}). Thus, adding noise on all angles simultaneously will result in cumulative coordinate displacements in later angle atom groups and may damage the pocket structure, which complicate the denoising process of the latter angles~\cite{zhang2023diffpack}. 

Compared with torsion angles without a natural order in molecules, the sidechain angles have a natural sequential order (shown in Fig. \ref{fig:3a}). To eliminate the cumulative effect and capture the pre- and post-dependencies when posing sidechains in complicated docking complex structures, we decompose the joint distribution of four sidechain torsional angles $\boldsymbol{\theta}_{1,2,3,4}^{sc}$ into individual conditional distributions:
\vspace{-0.5em}
\begin{equation}
\begin{small}
\begin{aligned}
\hspace{-0.5em}
p(\boldsymbol{\theta}^{sc}_{1,2,3,4})=p(\boldsymbol{\theta}^{sc}_1)\cdot p(\boldsymbol{\theta}^{sc}_2|\boldsymbol{\theta}^{sc}_1)\cdot p(\boldsymbol{\theta}^{sc}_3|\boldsymbol{\theta}^{sc}_{1,2})\cdot p(\boldsymbol{\theta}^{sc}_4|\boldsymbol{\theta}^{sc}_{1,2,3}).
\label{equ:8}
\end{aligned}
\end{small}
\end{equation}
This allows us to add noise to specific $\theta^{sc}_i$ angles in training iterations for simplifying denoising and generate the sidechain conformations step-by-step: we first predict $\boldsymbol{\theta}^{sc}_1$ for all residues based on the backbone and ligand structure. Then, using the complex with predicted $\boldsymbol{\theta}^{sc}_1$, we proceed to predict $\boldsymbol{\theta}^{sc}_2$, and so on for $\boldsymbol{\theta}^{sc}_3$ and $\boldsymbol{\theta}^{sc}_4$.

\noindent \textbf{Overall Framework.}
We train generative bridge models $\mathbb{P}^{\theta}_{\mathfrak{g}}$ on each geometry by minimizing $\mathcal{L}_\theta$ in Eq.\ref{equ:3} and obtain final prior bridge model $\mathbb{P}_\theta$ combining each bridge for probabilistic induced fitting. 

To parameterize the bridge $\mathbb{P}^\theta$, we use a relation-aware equivariant graph neural network on full atoms of complexes as a shared encoder and the equivariant pooling layer, invariant pooling layer for predicting SE(3)-equivariant (i.e., translations and rotations) and invariant (i.e., binding energy, torsion angles) quantities. Algorithmic description and more details on model architecture, data modeling, and training can be referred in Appendix. B.
\vspace{-0.5em}
\subsection{Injecting interaction prior into Geometry Bridges.}
\label{sec4.2}
Although we have defined the prior bridges and score matching loss on $\mathbb{G}$, we nevertheless develop the model and integrated interaction prior in 3D coordinates $\mathbf{X}$ directly. Providing the full 3D structure, rather than abstract elements of the product space, to the score model allows it to reason about physical interactions using \textit{SE(3)} equivariant model, and not be dependent on arbitrary definitions of torsion angles~\cite{jin2023unsupervised}. Further, interaction priors defined on abstract elements of geometries are less informative, and incorporating physical constraints defined on 3D coordinates can provide direct and accurate guidance over interactions within pocket-ligand complexes. Such design can also allow seamless integration of well-defined physical or statistical potential energy priors of protein-ligand interactions, which we choose in this paper. 

Bridging physical energy priors defined in 3D coordinates and prior bridges on geometry product space $\mathbb{G}$ is challenging as the prior drift terms in bridges are not directly comparable with the physical energy priors $E(\mathbf{X})$, nor the derivatives, i.e., physical forces $\mathbf{F} = \partial E/\partial{\mathbf{X}}$. To fill the gap and enjoy the benefits of both modeling schemes, we propose an \textit{Energy-to-Geometry mapping} method inspired by rigid body mechanics with Newton-Euler equations.

\begin{table*}[!htbp]
\begin{center}
\scriptsize
\begin{tabular}{@{}cccccccccccccc@{}}
\toprule
                     & \multicolumn{6}{c}{Holo Crystal Proteins}                                                                                 & \multicolumn{6}{c}{Apo ESMFold Proteins}                                                                                  &                                                                                 \\
                     & \multicolumn{3}{c}{Top-1 RMSD}                              & \multicolumn{3}{c}{Top-5 RMSD}                              & \multicolumn{3}{c}{Top-1 RMSD}                              & \multicolumn{3}{c}{Top-5 RMSD}                              & \multirow{2}{*}{\begin{tabular}[c]{@{}c@{}}Average \\ Runtime (s)\end{tabular}} \\
Method               & \%\textless{}2 & \%\textless{}5 & \multicolumn{1}{c|}{Med.} & \%\textless{}2 & \%\textless{}5 & \multicolumn{1}{c|}{Med.} & \%\textless{}2 & \%\textless{}5 & \multicolumn{1}{c|}{Med.} & \%\textless{}2 & \%\textless{}5 & \multicolumn{1}{c|}{Med.} &                                                                                 \\ \midrule
GNINA                & 34.6           & 62.6           & \multicolumn{1}{c|}{3.3}  & 46.8           & 78.1           & \multicolumn{1}{c|}{2.1}  & 9.9            & 31.7           & \multicolumn{1}{c|}{6.7}  & 18.1           & 53.5           & \multicolumn{1}{c|}{4.6}  & 251                                                                                \\
SMINA                & 21.5           & 50.1           & \multicolumn{1}{c|}{5.0}  & 37.2           & 70.8           & \multicolumn{1}{c|}{2.8}  & 4.9            & 19.2           & \multicolumn{1}{c|}{7.3}  & 9.6            & 43.4           & \multicolumn{1}{c|}{5.4}  & 246                                                                                \\
LeDock               & 24.0           & 48.6           & \multicolumn{1}{c|}{5.1}  & 39.4           & 69.8           & \multicolumn{1}{c|}{2.9}  & 10.9           & 36.7           & \multicolumn{1}{c|}{6.2}  & 19.5           & 59.0           & \multicolumn{1}{c|}{4.2}  & 7                                                                                \\
SurfLex              & 6.1            & 31.9           & \multicolumn{1}{c|}{6.7}  & 10.6           & 53.1           & \multicolumn{1}{c|}{4.7}  & 1.4            & 15.6           & \multicolumn{1}{c|}{7.4}  & 2.6            & 30.1           & \multicolumn{1}{c|}{6.1}  & 5                                                                                \\
QVina                & 21.3           & 50.9           & \multicolumn{1}{c|}{4.9}  & 36.8           & 71.8           & \multicolumn{1}{c|}{2.9}  & 5.0            & 19.5           & \multicolumn{1}{c|}{7.3}  & 9.9            & 42.9           & \multicolumn{1}{c|}{5.4}  & 2                                                                                \\ \midrule
GNINA-Flex           & 29.4           & 63.7           & \multicolumn{1}{c|}{3.5}  & 45.7           & 84.9           & \multicolumn{1}{c|}{2.2}  & 10.9           & 36.7           & \multicolumn{1}{c|}{6.2}  & 19.5           & 59.0           & \multicolumn{1}{c|}{4.2}  & 1072                                                                                \\
SMINA-Flex           & 20.5           & 46.2           & \multicolumn{1}{c|}{5.3}  & 34.4           & 73.3           & \multicolumn{1}{c|}{2.9}  & 5.8            & 18.4           & \multicolumn{1}{c|}{7.1}  & 11.1           & 42.0           & \multicolumn{1}{c|}{5.4}  & 963                                                                                \\
FlexPose             & 42.0           & 79.3           & \multicolumn{1}{c|}{2.2}  & -              & -              & \multicolumn{1}{c|}{-}    & 34.8           & \textbf{79.9}  & \multicolumn{1}{c|}{2.8}  & -              & -              & \multicolumn{1}{c|}{-}    & 12                                                                               \\
DiffDock(pocket, 40)     & 51.8           & 75.3           & \multicolumn{1}{c|}{2.0}  & 60.7           & 79.2           & \multicolumn{1}{c|}{1.9}  & 37.8           & 72.4           & \multicolumn{1}{c|}{2.6}  & 38.1           & 73.4           & \multicolumn{1}{c|}{2.5}  & 61                                                                              \\ \midrule
\textbf{Re-Dock (10)} & 51.9           & 77.4           & \multicolumn{1}{c|}{2.0}  & 62.2           & 83.6           & \multicolumn{1}{c|}{1.7}  & 39.0           & 74.8           & \multicolumn{1}{c|}{2.5}  & 39.0           & 74.8           & \multicolumn{1}{c|}{2.5}  & 15                                                                     \\
\textbf{Re-Dock (40)} & \textbf{53.9}  & \textbf{80.3}  & \multicolumn{1}{c|}{\textbf{1.8}}              & \textbf{65.0}  & \textbf{86.7}  & \multicolumn{1}{c|}{\textbf{1.4}}              & \textbf{42.9}  & 76.4           & \multicolumn{1}{c|}{\textbf{2.4}}              & \textbf{45.6}  & \textbf{78.2}  & \multicolumn{1}{c|}{\textbf{2.2}}              & 58                                                                              \\ \bottomrule
\end{tabular}
\end{center}
\vspace{-0.5em}
\caption{Performance of flexible redocking on the PBDBind test set (bound, holo-crystal proteins) and its corresponding ESMFold predicted Apo structures (unbound). The best metrics are marked by \textbf{bold}. In parenthesis, we specify the number of poses sampled from the generative model. It is worth noting that only our \texttt{Re-Dock} has no access to the holo-crystal proteins. Our \texttt{Re-Dock} surpasses all baselines across different metrics and settings with affordable inference time, which demonstrates our effectiveness and efficiency on flexible redocking as well as apo docking,  and the advantages of our geometric prior bridge with sidechain flexibility.}
\label{tab:1}
\vspace{-1em}
\end{table*}

\noindent \textbf{Energy-to-Geometry Mapping.}
There are inherent connections between energy and rigid body motions in rigid body mechanics. Given a potential energy $E(\mathbf{X})$, we can calculate the forces $\mathbf{F} = \partial E/\partial{\mathbf{X}}$ acting on a rigid body system. Then, the system's combined translational and rotational dynamics can be described in the Newton-Euler equations (Eq.~\ref{equ:10}). It is natural to take inspiration from rigid body mechanics to convert prior energy defined in 3D coordinates to corresponding geometric updates as our extra drift terms $f_t^{\mathfrak{g}}(Z_t)$ in geometric prior bridges.

Too short distances or clashes between atoms of ligands and pockets in the complex conformations generation process will induce abnormal Van der Waals forces. Thus, we incorporate anti-clash potential prior in our bridges. Following \cite{BortoliDSB2021,corso2023diffdock}, we choose \{$\boldsymbol{x} \in \mathbb{R}^3: S(\boldsymbol{x})=\gamma$\} where $S(\boldsymbol{x})=-\sigma\ln\left(\sum_{j=1}^{N_P}\exp\left(-\|\boldsymbol{x}-\boldsymbol{x}_P^{j}\|^2/\sigma\right)\right)$ as the descriptor of the protein surface. $N_P$ is the atom number of pockets and $\{\boldsymbol{x}_P^j\}_{j=1}^{N_P}$ is the set of pocket atom coordinate vectors. The anti-clash potential we used can be derived as follows:
\vspace{-0.5em}
\begin{equation}
\begin{small}
\begin{aligned}
\hspace{-0.5em}
E_{clash}(\boldsymbol{X}) = -\sum_i^{N_M}\max{(0,\gamma-S(\boldsymbol{x}_t^{(i)}))},
\label{equ:9}
\end{aligned}
\end{small}
\vspace{-0.5em}
\end{equation}
where $N_M$ is the atom number of molecules. We also include the \textit{Amber} physical interaction energy terms as prior potentials. The overall interaction potential energy prior $E(\boldsymbol{X})$ is a direct sum detailed in the Appendix. D.

\noindent \textbf{Newton-Euler Equation}
The Newton-Euler equations explain a rigid body’s combined translational and rotational dynamics. In the Center of Mass (CoM) frame, this can be written in matrix form as:
\begin{equation}
\begin{small}
\begin{aligned}
\hspace{-0.5em}
\begin{pmatrix}
\boldsymbol{F}\\\boldsymbol{\tau}\end{pmatrix}=\begin{pmatrix}m\mathbf{I}&0\\0&\mathbf{I}_c\end{pmatrix}\begin{pmatrix}d\boldsymbol{v}/dt\\d\boldsymbol{\omega}/dt\end{pmatrix}+\begin{pmatrix}0\\\boldsymbol{\omega}\times\mathbf{I}_c\boldsymbol{\omega}
\end{pmatrix}
\label{equ:10}
\end{aligned}
\end{small}
\end{equation}
where $\boldsymbol{F}$ and $\boldsymbol{\tau}$ are the total force and torque acting on CoM, $\boldsymbol{v}$ and $\boldsymbol{\omega}$ are the velocity of CoM and the angular velocity around CoM, $m$ and $\mathbf{I}_c$ is the mass and inertia matrix of the rigid body, which are constant for a given rigid body. 

In our \textit{Energy-to-Geometry mapping} module, the force $\boldsymbol{f}^i_t$ act on each ligand atom $i$ is defined as the gradient of the interaction energy prior function $\boldsymbol{f}^i_t = \partial E(\boldsymbol{X_t})/\partial\boldsymbol{x}_t^i$. We can then calculate the total force $\boldsymbol{F_t}$ and torque $\boldsymbol{\tau}_t$ for the ligand in the discrete-time step $t$:
\vspace{-0.5em}
\begin{equation}
\begin{small}
\boldsymbol{F}_t = \sum_{i\in \mathcal{V}_L}\boldsymbol{f}^i_t, \space \boldsymbol{\tau}_t = \sum_{i\in\mathcal{V}_L}(\boldsymbol{x}_t^i - \boldsymbol{p}^c_t) \times \boldsymbol{f}^i_t.
\label{equ:11}
\end{small}
\vspace{-0.5em}
\end{equation}
where $\boldsymbol{p}^c_t$ is the Center of Mass. We assume the system is stationary at each discrete time step $t$~\cite{guan2023linkernet, jin2023unsupervised}. Furthermore, we can specify changes of torsion angles to be \textit{disentangled} from rotations or translations~\cite{corso2023diffdock} to ensure torsional updates cause no linear or angular momentum, i.e., $\boldsymbol{\omega}_t = 0$ and $\boldsymbol{v}_t=0$. Thus, the Newton-Euler equations (Eq.\ref{equ:10}) can be simplified as $\boldsymbol{F}_t=m\frac{d\boldsymbol{v}_t}{dt}$ and $\boldsymbol{\tau}_t=\boldsymbol{I}_c\frac{d\boldsymbol{\omega}_t}{dt}$. For a short enough time period $\Delta t$, we have the velocity and angular velocity of the ligand as $\boldsymbol{\omega}_{t+\Delta t}=\boldsymbol{I}_c^{-1}\boldsymbol{\tau}_t\Delta t$ and $\boldsymbol{v}_{t+\Delta t}=\frac{1}{m}\boldsymbol{F}_t\Delta t$. Assuming each atom in the ligand has the unit mass and setting the time period $\Delta t$ as hyper-parameter $\alpha$, the prior drift terms in prior geometric bridges $\mathbb{Q}^{\mathfrak{g}}_{bb,f}$(Eq.~\ref{equ:5}) can be designed as follows:
\vspace{-0.5em}
\begin{equation}
\begin{small}
f_t^{\boldsymbol{r}}(Z_t) = \frac{\alpha}{|\mathcal{V}_L|}\boldsymbol{F}_t, f_t^{R}(Z_t) = \alpha\boldsymbol{I}_c^{-1}\boldsymbol{\tau}_t
\label{equ:12}
\end{small}
\vspace{-0.5em}
\end{equation}
where $|\mathcal{V}_L|$ denotes the number of atoms in the ligand. More details and analysis (e.g., the equivariance analysis and $f_t^{\boldsymbol{\theta}}$) can be referred to the Appendix. D.

\subsection{learning bridges by co-modeling energy and poses}
\label{sec4.3}
We have designed interaction-aware prior bridges on geometries via Eq.\ref{equ:5} and Eq.\ref{equ:12}. Next we will discuss how to better parameterize the learnable drift $s_t^\theta$ in Eq.\ref{equ:0} for learning the bridges. Specifically, we assume the learnable drift has a form of $s_t^{\theta}=\beta \hat{f_t} + \hat{s}_t^\theta$ where $\beta$ can be another learnable parameter, $\hat{s}_t^\theta$ is direct predicted scores and $\hat{f_t}$ is also calculated using \textit{Energy-to-Geometry mapping} with \textit{learnable} interaction energy $E^\theta(\boldsymbol{X}, \mathcal{G})$ as in equation \ref{equ:12}. 

This parameterization allows learning of prior drifts and the binding energy via an additional head at the same time. After training, $E^\theta(\boldsymbol{X}, \mathcal{G})$ can also serve as the \textit{confidence model}~\cite{corso2023diffdock} in addition to calculating $s_t^\theta$.

This idea has a connection with the Energy-Based model (EBMs)~\cite{jin2023unsupervised}, where the likelihood of a data point $p(\boldsymbol{X}, \mathcal{G}) \propto \exp(-E^\theta(\boldsymbol{X}, \mathcal{G}))$. Our training process can be seen as training
EBMs using denoising score matching with prior, and then we can interpret the learned energy as protein-ligand interaction energy, naturally including the interaction potential energy prior. 

Our method combines parameterizing the score directly (leaving the energy function \textit{implicit}) and \textit{explicitly} modeling the energy function. Thus, we enjoy both advantages of sampling quality and likelihood estimation to co-model energy and poses for imitating energy-constrained \textbf{induced-fit} process using denoising score matching loss in Eq.~\ref{equ:3}. 
\vspace{-1em}
\section{Experiments}
\vspace{-0.5em}
In this section, we justify the advantages of the proposed \texttt{Re-Dock} with comprehensive experiments. The experimental setup is introduced in Section \ref{sec5.1}. We aim to answer six research questions. \textbf{Q1:} How effective is \texttt{Re-Dock} for flexible re-docking and apo-docking tasks? \textbf{Q2:} Can \texttt{Re-Dock} generate accurate sidechain conformations? \textbf{Q3:} Can \texttt{Re-Dock} generalize well to more challenging and realistic tasks (e.g. cross-dock)? \textbf{Q4:} How do key framework designs impact the performance of \texttt{Re-Dock}? Can \texttt{Re-Dock} use fewer sampling steps to further speed up large-scale screening? \textbf{Q5:} Can \texttt{Re-Dock} generate physically valid samples? \textbf{Q6:} What's the quality of generated samples of \texttt{Re-Dock}? 
\begin{table*}[!htbp]
\begin{center}
\scriptsize
\begin{tabular}{@{}ccccccccccc@{}}
\toprule
                     & \multicolumn{5}{c}{Holo Crystal Proteins}                                                                                & \multicolumn{5}{c}{Apo ESMFold Proteins}                                                                \\
                     & \multicolumn{3}{c}{Top-1 SC-RMSD}                                   & \multicolumn{2}{c}{Top-5 SC-RMSD}                  & \multicolumn{3}{c}{Top-1 SC-RMSD}                                   & \multicolumn{2}{c}{Top-5 SC-RMSD} \\
Method               & \%\textless{}1 & \%\textless{}2 & \multicolumn{1}{c|}{Med.}         & \%\textless{}1 & \multicolumn{1}{c|}{Med.}         & \%\textless{}1 & \%\textless{}2 & \multicolumn{1}{c|}{Med.}         & \%\textless{}1   & Med.           \\ \midrule
GNINA-Flex           & 3.3            & 71.9           & \multicolumn{1}{c|}{1.7}          & 7.7            & \multicolumn{1}{c|}{1.4}          & 0.6            & 31.0           & \multicolumn{1}{c|}{2.5}          & 1.8              & 2.0            \\
SMINA-Flex           & 2.0            & 63.8           & \multicolumn{1}{c|}{1.8}          & 8.3            & \multicolumn{1}{c|}{1.4}          & 0.6            & 34.4           & \multicolumn{1}{c|}{2.4}          & 1.8              & 2.0            \\ \midrule
\textbf{Re-Dock (10)} & 86.8           & 95.4           & \multicolumn{1}{c|}{0.6}          & 93.1           & \multicolumn{1}{c|}{0.6}          & \textbf{39.8}  & 80.2           & \multicolumn{1}{c|}{1.2}          & 45.7             & 1.1            \\
\textbf{Re-Dock (40)} & \textbf{87.5}  & \textbf{100.0} & \multicolumn{1}{c|}{\textbf{0.6}} & \textbf{93.6}  & \multicolumn{1}{c|}{\textbf{0.6}} & 38.4           & \textbf{82.5}  & \multicolumn{1}{c|}{\textbf{1.2}} & \textbf{46.5}    & \textbf{1.1}   \\ \bottomrule
\end{tabular}
\end{center}
\vspace{-0.5em}
\caption{Performance of sidechain generation on PBDBind test set and its corresponding ESMFold predicted Apo structures. The best metrics are marked by \textbf{bold}. In parenthesis, we specify the number of poses sampled from the generative model. The results suggest our superior sidechain generation performance.}
\label{tab:2}
\vspace{-1em}
\end{table*}

\begin{table}[htbp]
\begin{center}
\scriptsize
\begin{tabular}{c|cc}
\hline
                       & \multicolumn{2}{c}{Top-1 RMSD} \\
Methods                & \%\textless{}2  & Med.         \\ \hline
DiffDock (pocket, 20 sampling steps, 40)              & 51.8            & 2.0          \\
Re-Dock (20 sampling steps, 40)                & \textbf{53.9}   & \textbf{1.8} \\ \hline
w/o Sidechain Generation & 39.8            & 3.3          \\
w/o Prior Energy        & 50.3            & 2.4          \\  \hline
DiffDock (pocket, interaction prior guided sampling, 40)  & 52.7            & 1.9          \\
Re-Dock (10 sampling steps, 40)      & 52.1            & 2.0          \\ \hline

\end{tabular}
\end{center}
\vspace{-1em}
\caption{Ablation study for designed components on flexible redocking task in PDBBind test set. The best metrics are marked by \textbf{bold}. In parenthesis, we specify the number of poses sampled from the generative model. w/o Sidechain Generation refers to removing the sidechain torsion sampling process in inference. w/o Prior Energy means training a vanilla geometric bridge without prior drift. Interaction prior guided sampling adds classifier guidance to DiffDock(pocket) using \textit{Energy-to-Geometry mapping}. }
\label{tab:4}
\vspace{-1.5em}
\end{table}

\vspace{-1em}
\subsection{Experimental Setups}
\vspace{-0.5em}
\label{sec5.1}
To construct a rigorous and reasonable benchmark for flexible docking, we design and set up four tasks with increasing modeling flexibility and difficulty. As mentioned above, we focus on results in the pocket-aware setting. Pockets are defined as residues within 8 \AA\  of ligands. Details and generalized results with pocket predictions are in the Appendix. 

\noindent \textbf{Evaluation Tasks.}
Tasks include:
a) \textbf{Flexible re-docking}, modified from standard re-docking tasks~\cite{corso2023diffdock, pei2023fabind}. Instead of using holo protein structures as input, we randomize pocket sidechain conformations at first and require the docking models to predict bound ligand poses based on noisy pockets. 
b) \textbf{Apo-dock}, which takes the unbound pocket structures as input. This task aligns with realistic scenarios, and we conduct it on the \textit{crystal apo protein structures} in addition to their \textit{computational alternatives} used in previous works~\cite{corso2023diffdock}.
c) \textbf{Sidechain pose prediction}. We also need to generate pocket sidechain poses to better understand interactions. Similar to the ligand pose prediction, we formulate this task to assess generated samples.
d) \textbf{Cross-dock}. It's an important real-world task for drug discovery~\cite{zhang2023learning}. The task involves predicting the ligand poses using pocket structures that are bound with different ligands, which can be biased and misleading. 
\begin{table*}[!htbp]
\begin{center}
\scriptsize
\begin{tabular}{@{}cccccccccc@{}}
\toprule
                     & \multicolumn{3}{c}{CrossDock}                                       & \multicolumn{6}{c}{Apo Crystal Proteins}                                                                             \\
                     & \multicolumn{3}{c}{Top-5 RMSD}                                      & \multicolumn{3}{c}{Top-1 RMSD}                                      & \multicolumn{3}{c}{Top-5 RMSD}                 \\
Method               & \%\textless{}2 & \%\textless{}5 & \multicolumn{1}{c|}{Med.}         & \%\textless{}2 & \%\textless{}5 & \multicolumn{1}{c|}{Med.}         & \%\textless{}2 & \%\textless{}5 & Med.         \\ \midrule
GNINA                & 23.5           & 75.0           & \multicolumn{1}{c|}{3.4}          & 7.0            & 20.7           & \multicolumn{1}{c|}{8.6}          & 8.9            & 29.6           & 6.9          \\
SMINA                & 16.8           & 72.2           & \multicolumn{1}{c|}{3.7}          & 5.6            & 15.3           & \multicolumn{1}{c|}{8.7}          & 8.4            & 27.3           & 7.2          \\
LeDock               & 10.3           & 65.3           & \multicolumn{1}{c|}{4.3}          & 3.3            & 12.6           & \multicolumn{1}{c|}{8.7}          & 5.9            & 27.5           & 6.7          \\
SurfLex              & 7.8            & 61.0           & \multicolumn{1}{c|}{4.4}          & 1.1            & 13.3           & \multicolumn{1}{c|}{8.0}          & 1.8            & 18.1           & 7.3          \\
QVina                & 17.3           & 72.2           & \multicolumn{1}{c|}{3.4}          & 6.4            & 16.0           & \multicolumn{1}{c|}{8.6}          & 9.2            & 28.4           & 7.3          \\ \midrule
GNINA-Flex           & 19.6           & 81.1           & \multicolumn{1}{c|}{3.0}          & 7.0            & 19.6           & \multicolumn{1}{c|}{8.3}          & 9.8            & 35.7           & 6.5          \\
SMINA-Flex           & 26.2           & 78.2           & \multicolumn{1}{c|}{3.1}          & 5.4            & 18.5           & \multicolumn{1}{c|}{8.2}          & 9.2            & 33.1           & 6.4          \\
$FlexPose^*$             & 13.8           & 77.4           & \multicolumn{1}{c|}{3.7}          & 23.3           & \textbf{61.7}  & \multicolumn{1}{c|}{3.9}          & -              & -              & -            \\
DiffDock(pocket)     & -              & -              & \multicolumn{1}{c|}{-}            & 30.3           & 50.6           & \multicolumn{1}{c|}{3.9}          & 33.7           & 51.2           & 3.7          \\ \midrule
\textbf{Re-Dock (10)} & \textbf{36.1}  & \textbf{91.8}  & \multicolumn{1}{c|}{\textbf{2.6}} & 31.3           & 51.7           & \multicolumn{1}{c|}{3.9}          & 35.6           & 54.9           & 3.6          \\
\textbf{Re-Dock (40)} & -     & -              & \multicolumn{1}{c|}{-}   & \textbf{34.4}  & 52.5           & \multicolumn{1}{c|}{\textbf{3.7}} & \textbf{38.6}  & \textbf{59.2}  & \textbf{3.3} \\ \bottomrule
\end{tabular}
\end{center}
\vspace{-1em}
\caption{Performance comparison on apo crystal re-docking and a more challenging task, cross-dock. The best metrics are marked by \textbf{bold} and we only generate 10 samples per complex for cross-dock to reduce computational burden. Methods with * only have Top-1 results. In addition to predicted approximated apo-structures, we provide promising results on apo crystal structures which are frequently used in real-world applications. Cross-dock requires the docking methods to predict correct ligand poses based on pocket structure bound with another ligand, which can be misleading and thus challenging. The results demonstrate the effectiveness and advantages of explicit modeling of flexible sidechains and our geometric prior bridge.}
\label{tab:3}
\vspace{-1em}
\end{table*}

\noindent \textbf{Datasets.}
We conduct training, flexible redocking, and sidechain pose prediction evaluation on the PDBBind v2020 dataset~\cite{liu2017PDBBind} with the time-based dataset split following previous works~\cite{corso2023diffdock, pei2023fabind}. For apo-docking, we use the ESMFold on the test set of PDBBind to obtain predicted apo structures. We also curate corresponding apo crystal structures with the PDBBind test via searching against Protein Data Bank~\cite{burley2021PDB}. For cross-dock, we collect complex structures bound with different ligands for 60 important drug targets and select 7 representative targets for evaluation, resulting in over 10000 complexes. 

\noindent \textbf{Metrics.}
We use Ligand RMSD as the evaluation metric, which calculates the root-mean-square deviation between the predicted and the holo crystal ligand atomic Cartesian coordinates. Following \cite{corso2023diffdock}, we report the percentage of predictions with \textit{RMSD} below 2 and 5, the median \texttt{RMSD} (Med.) and average runtime per complex. For sidechain pose prediction, we use a similar metric named SC-RMSD and report results with different thresholds of 1 and 2. For generated structures, the perfect match with the crystal structures is unrealistic, as they differ in bond lengths. To compensate for this fact, we use a relative measure that compares the SC-RMSD before and after the prediction.

\noindent \textbf{Baselines.}
We compare \texttt{Re-Dock} with state-of-the-art search-based methods SMINA~\cite{koes2013smina}, GNINA~\cite{mcnutt2021gnina}, LeDock~\cite{zhao2013LeDock}, SurfLex~\cite{jain2003surflex}, and Qvina~\cite{alhossary2015qvina}, and the recent deep learning method DiffDock~\cite{corso2023diffdock}. We implemented and retrained DiffDock(pocket) with pockets as input for comparison. We also include flexible docking baselines GNINA-Flex, SMINA-Flex, and FlexPose~\cite{dong2023FlexPose}. For generative models, we sample various poses per complex and take the most accurate pose out of 1 or 5 highest-ranked predictions (Top-1 or Top-5) according to the confidence model.

Extensive details about the experimental setup including data curation and training can be found in the Appendix. C.
\subsection{Results on flexible redocking and apo docking (Q1)}
In the context of flexible redocking, we observe the notable performance of our \texttt{Re-Dock}. Across all metrics, \texttt{Re-Dock} ranks as the best as presented in the left half of Table~\ref{tab:1}. It is worth noting that, all baselines in the flexible redocking task take the holo pocket structure as input for they are not trained for or perform much worse on noisy structures, while our \texttt{Re-Dock} receive randomized pocket sidechain poses which are more challenging. Detailed results can be referred in the Appendix. C. For Apo docking on both ESMFold~\cite{lin2022ESMFold} predicted structures and apo crystal structures, our \texttt{Re-Dock} also has leading performances as shown in the right half of Table~\ref{tab:1} and Table~\ref{tab:3}. The results suggest the advantage of our geometric prior bridge and explicit sidechain pose modeling.
\subsection{Quality of sidechain pose generation (Q2)}
Table~\ref{tab:2} shows the results of sidechain pose prediction in the PDBBind test set and corresponding apo structures. Our \texttt{Re-Dock} significantly outperforms all flexible docking baselines, especially in the metric of the percentage of SC-RMSD $<$ 1, highlighting the fine-grain modeling ability of our method. Different from the flexible redocking task above, all methods in the sidechain pose prediction task receive randomized pocket sidechain poses. 
\subsection{Performace comparison on cross-dock (Q3)}
We test our \texttt{Re-Dock} on the challenging cross-docking task as shown in Table~\ref{tab:3}. Our method has achieved state-of-the-art performance on the benchmark test set, suggesting the comparative ability of \texttt{Re-Dock} on flexible docking. Our scheme of co-modeling pocket sidechain and ligand flexibility in a probabilistic induced fitting process contributes to the promising generalization ability over realistic docking scenarios and real-world applications. Since the size of the dockings per pocket is unevenly distributed, we report the pocket-level cross-docking performance via a pocket-normalized score~\cite{brocidiacono2023plantain}.
\subsection{Ablation Study (Q4)}
As table~\ref{tab:4} suggests, components designed for \texttt{Re-Dock} all contribute to our superior performance. The results highlight the contribution of sidechain generation and prior energy. Notably, modifying the training process with a prior bridge yields better results than using classifier guidance~\cite{dhariwal2021diffusion} in sampling. Further, we achieve better results than diffusion with fewer time steps.
\subsection{More results on the PoseBuster Benchmark (Q5)}
Table~\ref{tab:5} reports more results on PoseBuster~\cite{buttenschoen2024posebusters} benchmark. We achieve the best performance over deep learning-based methods especially in the PB-Valid check setting, demonstrating the superior ability of geometric prior bridge for generating realistic samples.
\begin{table}[htbp]
\begin{center}
\scriptsize
\begin{tabular}{c|cc}
\hline
                & \multicolumn{2}{c}{Top-1 RMSD}            \\
Methods         & \%\textless{}2 & \%\textless{}2\&PB-Valid \\ \hline
Gold~\cite{verdonk2003gold}            & 58.0           & 55.0                     \\
Vina~\cite{autodock2021}            & 60.0           & 58.0                     \\ \hline
DeepDock~\cite{liao2019deepdock}        & 20.0           & 5.2                      \\
Uni-Mol~\cite{zhou2023unimol}         & 22.0           & 2.0                      \\
TankBind~\cite{Lu2022TankBind}        & 16.0           & 3.3                      \\
DiffDock(40)       & 38.0           & 12.0                     \\ \hline
\textbf{Re-Dock}(40) & \textbf{50.7}  & \textbf{32.8}            \\ \hline
\end{tabular}
\end{center}
\vspace{-1em}
\caption{Flexible redocking results on PoseBuster benchmark. PB-Valid means the generated docking samples pass the PoseBuster Valid check and are physically valid~\cite{buttenschoen2024posebusters}. The best results among deep learning-based methods are \textbf{bold}. Results demonstrate the effectiveness of \texttt{Re-Dock} for generating realistic (i.e., physically valid) samples.}
\label{tab:5}
\vspace{-1em}
\end{table}
\vspace{-0.5em}
\subsection{Case Study (Q6)}
\begin{figure}[!tbp]
    \begin{center}
        \includegraphics[width=1.0\columnwidth]{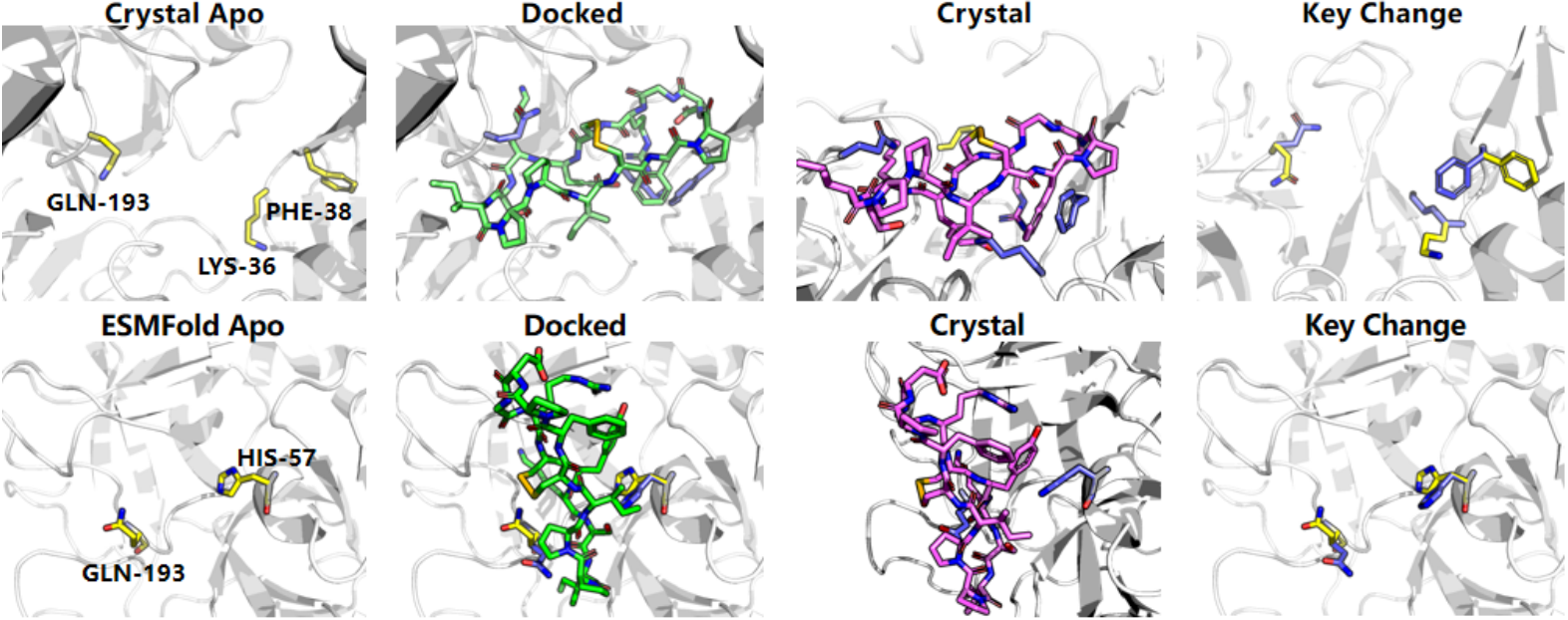}
    \end{center}
    \vspace{-1em}
    \caption{Case Studies on Plasmin~\cite{gallus1976prevention}, a real-world protein target and its inhibitor ligand SFTI-1~\cite{de2021sunflower}. The numbers (e.g. HIS-57) indicate the name and id of residues. \texttt{Re-Dock} replicates the bound crystal conformations for ligands (Docked) and sidechains (Key Change) on the apo (Top, PDB: 1QRZ) and predicted (Bottom) structure of plasmin. }
    \label{fig:4}
    \vspace{-1.5em}
\end{figure}

\vspace{-0.5em}
\noindent \textbf{\texttt{Re-Dock} can generate `induced' sidechain poses in realistic scenarios.} The first and last column of Fig.\ref{fig:4} shows \texttt{Re-Dock} can prompt the inward folding of residues that facilitate the accurate positioning of the SFTI-1 ligand in realistic scenarios using crystal apo or predicted structures. 

\noindent \textbf{\texttt{Re-Dock} generates valid ligand pose for real-world targets}
In the two middle columns of Fig.\ref{fig:4}, the real-world target Plasmin (PDB: 1QRZ) is not seen during training, but we can successfully generalize to this real-world challenge, predicting the accurate binding poses.

\vspace{-0.5em}
\section{Conclusion}
\vspace{-0.5em}
We introduce an under-explored docking task, flexible docking, which better aligns with realistic applications. Tailored to this task, we present \texttt{Re-Dock}, a geometric prior bridge generative model with the energy-to-geometry mapping inspired by the Newton-Euler equation in mechanics. We extend the prior bridge to the geometric manifold and propose a novel and general module that bridges the explicit interaction energy with implicit geometries to design geometric prior drifts and co-model the energy and poses. With explicit generative modeling of ligands and pocket sidechain poses, \texttt{Re-Dock} simulates the induced fitting process with a probabilistic one. The results on the proposed thoughtful benchmarks demonstrate our efficiency and effectiveness. Limitations still exist, including insufficient exploration of pocket prediction and end-to-end blind docking.

\section*{Impact Statements}
Unlike previous methods, \texttt{Re-Dock} is designed and trained for flexible docking and thus better generalizes to realistic applications with real-world drug targets and realistic results. It integrates binding energy prediction and can offer great value for many realistic drug discovery and protein engineering pipelines as well as disease understanding. With a better trade-off of efficiency and effectiveness (for instance, choosing a proper number of samples generated per complex and sampling steps would help), \texttt{Re-Dock} can be applied for both large-scale virtual screening and deeper analysis of protein-ligand interaction. While there exists the potential risk that docking methods could be misused to develop harmful drugs, it's important to note that drug development is subject to stringent oversight globally. This rigorous regulatory environment ensures that such misuses can be effectively managed and controlled.
\bibliography{Re-Dock}
\bibliographystyle{icml2023}

\newpage
\appendix
\onecolumn
\section{More related works}

\noindent \textbf{Protein Protein Docking} Protein-protein docking aims to predict the combined structure of two proteins from their individual shapes. This process, similar to molecular docking, typically assumes that proteins do not change shape when they bind, limiting their movements to rotations and translations in three dimensions. Searching-based protein-protein docking methods often start with a basic initialization and refine it gradually~\cite{yan2017hdock, chen2003zdock, de2010haddock}. Furthermore, some use template-based modeling to align a target protein with known structures~\cite{vakser2014protein}. Deep learning techniques~\cite{wu2023extracting, lin2023nonequispaced, wu2023quantifying, Hu2023complete, wu2022automated, zheng2023mmdesign} in this field are divided into two main types: one aims for immediate predictions, while the other focuses on refining predictions step by step. Research like ~\cite{ganea2021independent} targets predicting the precise adjustments needed for binding. Another study integrates physical principles into an energy-based model to predict the 3D structure of protein complexes~\cite{sverrisson2022physics}. Techniques that refine their predictions progressively, such as ALPHAFOLD-MULTIMER~\cite{evans2021protein}, fold multiple proteins together based on their sequences and related proteins. In parallel with this work, DOCKGPT~\cite{mcpartlon2023deep} offers a novel approach for precise and adaptable protein docking.

\noindent \textbf{Flexible Docking}
\textit{Flexible docking} is an open challenge for developing docking methods as it includes much more degrees of freedom than rigid docking. Methods often adopt effective assumptions for restricting the degrees of freedom, such as only considering the flexible sidechain conformations~\cite{alhossary2015qvina, taylor2003fds, hartmann2009docking}. ~\cite{zhang2023diffpack} develop generative models for sidechain packing (that's, without ligands), and ~\cite{plainer2023diffdock_pocket} explores pocket-specific docking with receptor sidechain flexibility. Different from previous diffusion-based methods for sidechain flexibility, we explore a novel generative framework, incorporate interaction prior for sufficient protein-ligand interaction modeling, and co-model the poses and binding energy. Without sidechain matching and additional training on ESMFold-generated structures, we achieve comparable or better results than \cite{plainer2023diffdock_pocket}.

\section{Model Architecture and Training}
\subsection{Model Architecture}
\noindent \textbf{Notations.}
For each protein-ligand complex, we represent it as a heterogeneous graph with both coarse and fine grain modeling $\mathcal{G}=(\mathcal{V}:=\{\mathcal{V}_L,\mathcal{V}_P^{res},\mathcal{V}_P^{atom}\},\mathcal{E}:=\{\mathcal{E}_{in}, \mathcal{E}_{ex}\})$. Here, the components $\mathcal{V}_L, \mathcal{V}_P^{atom}$ correspond to the nodes (i.e., the heavy atoms) of the ligand and protein (atom-level). $\mathcal{V}_P^{res}$ represent the nodes (i.e., the residues) of the protein (residue-level); $\mathcal{E}_{in}, \mathcal{E}_{ex}$ separately contain internal edges within each component and external edges across components. To be specific, each node in $\mathcal{V}$, i.e., $v_i = (h_i, x_i)$ is represented as a node embedding vector $h_i$ initialized with trainable type embedding vector and ESM-2 embedding~\cite{lin2022language} following \cite{corso2023diffdock}, and a coordinate vector $x_i \in \mathbb{R}^3$ of the corresponding (heavy) atoms (atom-level) or the alpha carbon of the residue backbone (residue-level). To be aware of the general gesture of ligand for better generation of the binding pose, we add a global node that connects to all atoms in the component coordinated at the mean of all coordinates.

Each edge in $\mathcal{E}$, i.e., $e_{ij}=(h_{ij}, x_{ij})$ consists of an edge embedding of node $(i,j)$ initialized with radial basis embedding of edge length and edge vector $x_{ij} = x_j - x_i$. For explicit and accurate modeling of protein-ligand interactions, we separate internal and external interactions ($\mathcal{E}_{in}$ and $\mathcal{E}_{ex}$ respectively) because of their different role in binding and distance scales. Furthermore, to model interactions in a fine-grained manner, We connect nodes using cutoffs dependent on the type of nodes they are connecting and assign the edges distinct edge types corresponding to different graph convolution kernels following \cite{zhang2023protein, kong2023conditional, corso2023diffdock}. We further denote the pocket-ligand subgraph as $\mathcal{G}_{p*}=(\mathcal{V}:=\{\mathcal{V}_L,\mathcal{V}_{p*}^{res},\mathcal{V}_{p*}^{atom}\},\mathcal{E}:=\{\mathcal{E}_{in}, \mathcal{E}_{ex}\})$ which is the input of the docking model.

\noindent \textbf{Architecture.}
We construct a heterogeneous graph with atoms and residue levels to reason the pocket-ligand interaction in a more fine-grained manner~\cite{zhang2023protein}. The edge constructions have considered various interaction types, including internal interactions ( atom-atom interactions, chemical bone interactions, residue-residue spatial and sequential interactions, atom-residue subordination relationship, etc.) and external interactions (pocket-ligand atom interactions, residue-atom interactions, framework-pocket interactions, etc.). We assign different graph convolution kernels to each edge type and model every interaction separately. These relation-aware schemes can enhance the modeling ability for accurate pocket-ligand interactions. 

The output of the score model must be in the tangent space $T_{\mathbf{r}}\mathbb{T}_{3}\oplus T_{R}SO(3)\oplus T_{\boldsymbol{\theta}}SO(2)^{m}$ and the predicted energy must be in $SE(3)$-invariant. The space $T_{\boldsymbol{r}}\mathbb{T}_3$ and $T_RSO(3)$ represent the translation and rotation (Euler) vectors which are both $SE(3)$-equivariant and $T_{\boldsymbol{\theta}}SO(2)^m$ corresponds to scores on $SE(3)$-invariant quantities (torsion angles).
To achieve such desire, we apply a $SE(3)$-equivariant convolution network(6 layers of graph convolutions on the heterogeneous graph implemented with e3nn library~\cite{geiger2022e3nn, corso2023diffdock}) as the shared encoder with equivariant and invariant pooling as heads as well as pseudo torque convolution~\cite{jing2022torsional} for predicting torsional updates. 

\subsection{Model training and implementation details}
Following \cite{corso2023diffdock}, we train and evaluate all the models on PDBBind~\cite{liu2017PDBBind} based on time-split dataset partition. We also 
adopt a conformer matching procedure described in \cite{jing2022torsional} for eliminating the distributional shift between RDKit-initialization used in inference and ground-truth ligand pose in the dataset. We summarize the training as algorithmic description as in Algorithm~\ref{alg:1}.

\begin{algorithm}[!htbp]
\caption{Learning diffusion generative models}
\label{alg:algorithm}
\textbf{Input}: Training pairs $\{(x^*,y \}$ where $x^*$ is the ground truth ligand pose and $y$ is the ground truth protein structure, RDKit preditions $\{c\}$, variance for each geometry $\{\sigma_\mathfrak{g}\}$, prior interaction energy $E(x,y)$ and $\mathbb{Q}^x_{\mathfrak{g}}$ the bridge in Eq.~\ref{equ:5}, and energy-to-geometry mapping $F$ and a diffusion model $\mathbb{P}^\theta$.\\
\textbf{Output}: Parameters $\theta$ of $\mathbb{P}_\theta$
\begin{algorithmic}[1] 
\STATE Randomly initialize the parameters of $\mathbb{P}_\theta$.
\FOR{$c, x^*, y \in \{(x^*, y, c)\} $}
\STATE Let $x_0 \leftarrow conformer\_align(x^*, c)$;
\STATE Sample $t \in Uni([0,1])$, $chi\_id sc \in {0,1,2,3}$;
\STATE Sample $\{\Delta \mathfrak{g}\}$ from diffusion kernels $\{p_t^\mathfrak{g}(\cdot|\sigma_\mathfrak{g})\}$;
\STATE Compute $(x_t, y_t) \leftarrow Apply(\{\Delta \mathfrak{g}\}, x_0, y, sc)$ and ${f_t^\mathfrak{g}} = {F(E(x_t, y_t))}$;
\STATE Predict scores;
\STATE Take optimization step on the sum of denoising score matching loss Eq.~\ref{equ:3} on each geometry;
\ENDFOR
\STATE \textbf{return} Diffusion Bridge $\mathbb{P}_\theta$.
\end{algorithmic}
\label{alg:1}
\end{algorithm}

\noindent \textbf{training detail.} We use Adam as optimizer with learning\_rate= 0.001 and exponential moving average of the weights during training, which we will use in inference. we update the moving average after every optimization step with a decay factor of 0.999. The batch size is 64. We run inference with 20 denoising steps on 500 validation complexes every 10 epochs and use the set of weights with the highest percentage of RMSDs less than 2\AA\  as the final score model. All baselines and our approach are implemented using the PyTorch 1.6.0 library with Intel(R) Xeon(R)Gold6240R@2.40GHz CPU and NVIDIA A100 GPU. We train our score model for 600 epochs (around 7 days). As we co-model the binding energy and poses, we don't need additional training for the confidence model as in \cite{corso2023diffdock}, but the confidence model can still be adopted in our framework for ranking. For inference, only a single GPU is required but we also implement parallel inference in multi-gpu settings for large scale screening task like cross-dock (around 10000 complexes, 10 poses per complexes and 20 time steps).

\noindent \textbf{Runtime} Similar to previous methods, we exclude the data preprocessing time and perform runtime analysis on single cpu core (plus additional 
gpus for methods that can utilize gpus). Preprocessing mainly consists of a forward
pass of ESM2 to generate the protein language model embeddings, RDKit’s conformer generation, and the conversion of the protein into a radius graph. We measured the inference time when running on the same device for training.

\section{Experimental setup}
To evaluate the performance of our method, we design a systematic benchmark mimicking the flexible and realistic setting in drug discovery pipelines, including redocking, unbound(apo)-structure docking (including predicted structures and experimental structures), and cross-dock (docking to bound structure with another ligand, which is very challenging but common in drug discovery) experiments in the PDBBind dataset. Re-Dock has shown competitive results across all tasks in various metrics assessing conformational plausibility. This suggests our model can handle the flexibility of both pockets and ligands and has the potential to be helpful in realistic drug discovery pipelines. 
\subsection{Data Curation}
\noindent \textbf{Crystal apo structure for PDBBind time split test set.} The APO structures are retrieved from the Protein Data Bank (PDB)~\cite{burley2021PDB} by first extracting the sequence of HOLO structures and conducting a BLAST~\cite{ye2006blast} search against the PDB database. Each hit protein is then structurally aligned to the holo-structure using PyMOL~\cite{delano2002pymol}, focusing on the superposition of corresponding $C_\alpha$ atoms of amino acid residues. Post alignment, structures are assessed for quality and relevance similar to ApoBind~\cite{aggarwal2021apobind}: those with a backbone $C_\alpha$ Root Mean Square Deviation (RMSD) exceeding 15 Å, or those showing less than 80\% sequence identity or coverage compared to the full protein sequence, are rejected. Additionally, any hit structure with ligands located within 4 Å of any atoms in the crystal structure pose of the complex is also excluded.

\noindent \textbf{Crystal CrossDock structures.} The targets are from DUD-E~\cite{mysinger2012directory}, a dataset designed for the unbiased virtual screening task. Seven targets are selected, AKT1, AMPC, CXCR4, GCR, HIVPR, HIVRT, and KIF11, which are representatives of kinase, other enzymes, G protein-coupled receptors, nuclear receptors, protease, other enzymes, and miscellaneous classes. Within each target, we retrieve different protein-ligand structures from PDB using a similar method and conditions of the construction of crystal apo time split test set.

For search-based method, we use their official software suits and for deep-learning-based baselines, we use their official implementation and weights. To develop diffdock (pocket), we adopt the same pocket truncation method as our \textsc{Re-Dock} and full-atom graph representation as it used for their confidence model. Other settings keep the same with the original diffdock.

\section{More details about the bridge}
The space of the pocket-ligand poses $\mathbb{R}^{3(m+n)}$ (m and n are the number of pocket sidechain and ligand atoms) is huge and encompasses far more degrees of freedom than are relevant in molecular docking. The complex flexibility lies almost entirely in the torsion angles at rotatable bonds~\cite{corso2023diffdock}; thus, we incorporate geometric prior, which is known in advance (e.g., fixed bond lengths, angles, and essentially rigid small rings) with a seed (randomized only in torsion angles) pocket sidechain and ligand conformations \textbf{C} in isolation, and model the pocket-ligand complex pose in an $(m+6)$-dimensional submanifold $\mathcal{M}_C$, where $m$ is the number of rotatable bonds and six additional degrees of freedom come from rototranslations relative to the fixed protein backbone.

While diffusion generative models have been applied to molecular docking, existing approaches are ill-suited for flexible and realistic docking scenarios, where we need to co-model the flexibility of pocket sidechains and ligands as well as their interactions. To develop Re-Dock, we recognize the dynamic and interactive nature of docking with induced fit; thus, we use the diffusion bridge model to inject physical priors of molecular interactions into the generative process. To simplify the modeling process and incorporate geometric prior (e.g. fixed bond lengths, bond angles between atoms), we follow the successful experience in the field of conformational generation and develop geometry-based generative models.

We construct such a bridge with prior of the atom interaction potentials regarding how the diffusion process should look like for generating each given data point over the product space of geometries: global rotation/translation of ligands and local molecular torsion/residue sidechain angles.

As the induced-fit process is driven by the inherent interaction energies and generative docking needs a confidence model for ranking predicted poses~\cite{corso2023diffdock}, we include learnable interaction energy with an additional energy head to co-model the interaction energy and conformation.

We use prior energy terms in Amber~\cite{case2021amber}, including:
1. The Lennard-Jones (LJ) energy $E_{LJ}(x)=\sum_{i\neq j}e(\begin{Vmatrix}x_{i}^{r}-x_{j}^{r}\end{Vmatrix})\mathrm{~and~}e(\ell)=(\sigma/\ell)^{12} - 2(\sigma/\ell)^6$. The parameter $\sigma$ is an approximation for average nucleus distance. 2. The nuclei-nuclei repulsion (Coulomb) electromagnetic potential energy is $E_{Coulomb}=\kappa\sum_{ij}q(\hat{x}_{i}^{h})q(\hat{x}_{j}^{h})/\left\|x_{i}^{r}-x_{j}^{r}\right\|$, where $\kappa$ is Coulomb constant and $q(r)$ denotes the point charge of atom of type $r$, which depends on the number of protons. We use these energies based on the same threhold with the $E_{clash}$ in Equation.~\ref{equ:9} as the ligand and sidechains are close enough.

As the \textit{energy-to-geometry mappp} is derived from rigid-body mechanics, it is naturally equivariant to roto-translations (i.e. rotations and translations). As the calculation of prior drift on torsion manifold $SO(2)^M$, where $M=m_{lig}+m_{sc}$ is the number of rotatable bonds, is complex and requires extensive computations, we omit it for simplicity. The torsion force can be described as the difference between the torques applied on each side of the rotatable bonds,but its adaptation with $SO(2)^M$ manifold and autoregressive sidechain torsion update is non-trivial and we leave it as future work.
\section{More Results}
We provide generlized results with predicted pocket with p2rank~\cite{krivak2018p2rank} in table~\ref{app_tab:1}
\begin{table*}[!htbp]
\begin{center}
\scriptsize
\begin{tabular}{@{}cccccccccccccc@{}}
\toprule
                     & \multicolumn{6}{c}{Holo Crystal Proteins}                                                                                 & \multicolumn{6}{c}{Apo ESMFold Proteins}                                                                                  &                                                                                 \\
                     & \multicolumn{3}{c}{Top-1 RMSD}                              & \multicolumn{3}{c}{Top-5 RMSD}                              & \multicolumn{3}{c}{Top-1 RMSD}                              & \multicolumn{3}{c}{Top-5 RMSD}                              & \multirow{2}{*}{\begin{tabular}[c]{@{}c@{}}Average \\ Runtime (s)\end{tabular}} \\
Method               & \%\textless{}2 & \%\textless{}5 & \multicolumn{1}{c|}{Med.} & \%\textless{}2 & \%\textless{}5 & \multicolumn{1}{c|}{Med.} & \%\textless{}2 & \%\textless{}5 & \multicolumn{1}{c|}{Med.} & \%\textless{}2 & \%\textless{}5 & \multicolumn{1}{c|}{Med.} &                                                                                 \\ \midrule
GNINA                & 34.6           & 62.6           & \multicolumn{1}{c|}{3.3}  & 46.8           & 78.1           & \multicolumn{1}{c|}{2.1}  & 9.9            & 31.7           & \multicolumn{1}{c|}{6.7}  & 18.1           & 53.5           & \multicolumn{1}{c|}{4.6}  & 251                                                                                \\
SMINA                & 21.5           & 50.1           & \multicolumn{1}{c|}{5.0}  & 37.2           & 70.8           & \multicolumn{1}{c|}{2.8}  & 4.9            & 19.2           & \multicolumn{1}{c|}{7.3}  & 9.6            & 43.4           & \multicolumn{1}{c|}{5.4}  & 246                                                                                \\
LeDock               & 24.0           & 48.6           & \multicolumn{1}{c|}{5.1}  & 39.4           & 69.8           & \multicolumn{1}{c|}{2.9}  & 10.9           & 36.7           & \multicolumn{1}{c|}{6.2}  & 19.5           & 59.0           & \multicolumn{1}{c|}{4.2}  & 7                                                                                \\
SurfLex              & 6.1            & 31.9           & \multicolumn{1}{c|}{6.7}  & 10.6           & 53.1           & \multicolumn{1}{c|}{4.7}  & 1.4            & 15.6           & \multicolumn{1}{c|}{7.4}  & 2.6            & 30.1           & \multicolumn{1}{c|}{6.1}  & 5                                                                                \\
QVina                & 21.3           & 50.9           & \multicolumn{1}{c|}{4.9}  & 36.8           & 71.8           & \multicolumn{1}{c|}{2.9}  & 5.0            & 19.5           & \multicolumn{1}{c|}{7.3}  & 9.9            & 42.9           & \multicolumn{1}{c|}{5.4}  & 2                                                                                \\ \midrule
GNINA-Flex           & 29.4           & 63.7           & \multicolumn{1}{c|}{3.5}  & 45.7           & 84.9           & \multicolumn{1}{c|}{2.2}  & 10.9           & 36.7           & \multicolumn{1}{c|}{6.2}  & 19.5           & 59.0           & \multicolumn{1}{c|}{4.2}  & 1072                                                                                \\
SMINA-Flex           & 20.5           & 46.2           & \multicolumn{1}{c|}{5.3}  & 34.4           & 73.3           & \multicolumn{1}{c|}{2.9}  & 5.8            & 18.4           & \multicolumn{1}{c|}{7.1}  & 11.1           & 42.0           & \multicolumn{1}{c|}{5.4}  & 963                                                                                \\
FlexPose             & 42.0           & 79.3           & \multicolumn{1}{c|}{2.2}  & -              & -              & \multicolumn{1}{c|}{-}    & 34.8           & \textbf{79.9}  & \multicolumn{1}{c|}{2.8}  & -              & -              & \multicolumn{1}{c|}{-}    & 12                                                                               \\
DiffDock(pocket, 40)     & 51.8           & 75.3           & \multicolumn{1}{c|}{2.0}  & 60.7           & 79.2           & \multicolumn{1}{c|}{1.9}  & 37.8           & 72.4           & \multicolumn{1}{c|}{2.6}  & 38.1           & 73.4           & \multicolumn{1}{c|}{2.5}  & 61                                                                              \\ \midrule
\textbf{ReDock (10)} & 51.9           & 77.4           & \multicolumn{1}{c|}{2.0}  & 62.2           & 83.6           & \multicolumn{1}{c|}{1.7}  & 39.0           & 74.8           & \multicolumn{1}{c|}{2.5}  & 39.0           & 74.8           & \multicolumn{1}{c|}{2.5}  & 15                                                                     \\
\textbf{ReDock (40)} & \textbf{53.9}  & \textbf{80.3}  & \multicolumn{1}{c|}{\textbf{1.8}}              & \textbf{65.0}  & \textbf{86.7}  & \multicolumn{1}{c|}{\textbf{1.4}}              & \textbf{42.9}  & 76.4           & \multicolumn{1}{c|}{\textbf{2.4}}              & \textbf{45.6}  & \textbf{78.2}  & \multicolumn{1}{c|}{\textbf{2.2}}              & 58 \\
\textbf{ReDock (p2rank, 40)} & 40.2  &  65.7 & \multicolumn{1}{c|}{3.1}              & 45.6  & 77.8  & \multicolumn{1}{c|}{2.2}              & 22.8  & 56.7           & \multicolumn{1}{c|}{4.9}              & 32.8  & 65.2  & \multicolumn{1}{c|}{3.3}              & 58
\\ \bottomrule
\end{tabular}
\end{center}
\vspace{-0.5em}
\caption{Performance of flexible redocking on the PBDBind test set (bound, holo-crystal proteins) and its corresponding ESMFold predicted Apo structures (unbound). The best metrics are marked by \textbf{bold}. In parenthesis, we specify the number of poses sampled from the generative model. It is worth noting that only our \texttt{Re-Dock} has no access to the holo-crystal proteins. p2rank denotes we use the predicted pocket center with p2rank and a pocket radius of 15 \AA}
\label{app_tab:1}
\vspace{-1em}
\end{table*}



\end{document}